\newcommand\D {{\cal D}}
\documentstyle[12pt]{article}
\begin{document}
\pagestyle{empty}
\voffset=-2cm
\hoffset=-1cm
\textwidth=16cm
\textheight=22cm
\rightline{UG-10/94}
\rightline{FIAN/TD/17/94}
\rightline{hep-th/9411093}
\rightline{November 1994}
\vspace{1.8truecm}
\centerline{\bf  The Calogero Model and the Virasoro Symmetry}
\vspace{1.8truecm}
\centerline{\bf E.~Bergshoeff}
\vspace{.5truecm}
\centerline{Institute for Theoretical Physics}
\centerline{Nijenborgh 4, 9747 AG Groningen}
\centerline{The Netherlands}
\vspace{.5truecm}
\centerline{and}
\vspace{.5truecm}
\centerline{\bf M.~Vasiliev\footnote{
Supported in part by the Russian Fund of Fundamental Research,
grant N67123016.}
}
\vspace{.5truecm}
\centerline{I.E.~Tamm Theoretical Department, P.N.~Lebedev Physical Institute}
\centerline{117924, Leninsky Prospect 53, Moscow}
\centerline{Russia}
\vspace{1.5truecm}
\centerline{ABSTRACT}
\vspace{.5truecm}
We construct new realizations of the Virasoro algebra inspired by
the Calogero model. The Virasoro algebra we find acts as a kind of
spectrum-generating algebra of the Calogero model. We furthermore
present the superextension of these results and introduce a
class of higher-spin extensions of the Virasoro algebra which are
of the $W_\infty$ - type.

\vfill\eject
\pagestyle{plain}
\section{Introduction}
%\noindent{\bf 1. Introduction}

\bigskip

Over the years there has been an increasing interest in the possible
relationships between integrable systems and conformal field theory.
A well-known and well studied example is provided by the Liouville
model and, more generally, the Toda models which play a crucial role
in the study of non-critical string theories.

Another example of an integrable system
is the Calogero model which has been studied
intensely since its construction in 1969 \cite{Ca1,Ca2}. Recently,
further progress has been obtained in understanding the $N$-body
Calogero model
with harmonic interaction \cite{Po1,Br1}. In \cite{H,Po2,Br3} it was
argued that the Calogero model describes one-dimensional
reductions of anyonic systems \cite{ML,W}, e.g.~anyons at the
lowest Landau level in a strong magnetic field. In its turn,
anyon physics plays an
important role in understanding the (fractional) quantum Hall effect
(QHE)\cite{Hal} that assumes interesting links between the latter
and the Calogero model.

The recent
progress \cite{Po1,Br1} in the understanding of the (rational) Calogero
model was based
on an extension of the Heisenberg algebra, the so-called $SH_N(\nu)$ algebra
\cite{Br2}. For $N=1$ the algebra reduces to
the ordinary Heisenberg algebra which
underlies the higher-spin algebras in higher dimensions \cite{V}
as well as the standard realization of the Virasoro algebra
as vector
fields and of the $W_{1+\infty}$ algebra \cite{Pope}
as differential operators on
the circle.

The question we address in this paper is whether the
general $SH_N(\nu)$ algebra can lead to new realizations of the (super)
Virasoro algebra.
We show that this is indeed the case and, as a result, one discovers a
new interesting class of $W_\infty$ - type algebras containing higher-spin
currents in addition to the Virasoro generators. We expect that our results
will lead to
a better insight into possible relations between the Calogero
model and conformal field theory. Our results may also have
applications to the quantum Hall effect. Actually, it was
observed recently \cite{WH1} that the relevant approach to the QHE is
based on the
representation theory of the $W_{1+\infty}$ algebra developed
in \cite{Kac}. Since, on the other
hand, the algebras
considered in the present paper are responsible for the $N$ - body
excitations in the Calogero model and can be regarded as some
(extensions $\times$ deformations) of the ordinary $W_{1+\infty}$
algebra one can speculate that they might be relevant for the analysis
of the many-body excitations in the QHE.
\vspace{.5cm}
\section{The $\,SH_N(\nu)\,$ realization of the Virasoro Algebra}
%\noindent{\bf 2. The $\,SH_N(\nu)\,$ realization of the Virasoro
%Algebra}

\bigskip

Our starting point is the $S_N$-extended Heisenberg algebra
$SH_N(\nu)$ \cite{Po1,Br1,Br2} which can
be regarded as the algebra formed by the generating elements
$a_i, a_i^\dagger$ and $K_{ij} \quad (i,j = 1,...,N)$
obeying the following relations\footnote{
In this paper repeated indices do not imply summation.}:

\begin{eqnarray}
\label{eq:algebra}
[a^{(\dagger)}_i ,a^{(\dagger)}_j ]&=&0 \,,\quad \,
[a_i ,a^{\dagger}_j ]
= A_{ij} \equiv \delta_{ij }(1+\nu\sum_{l=1}^N K_{il})-\nu K_{ij}\,, \\
K_{ij}K_{jl}&=&K_{jl}K_{il}=K_{il}K_{ij}\,, \quad
{\rm for \ all}\ i \ne j, i \ne l, j \ne l\,, \\
(K_{ij})^2&=&I\,,\qquad K_{ij}=K_{ji}\,, \\
K_{ij}K_{mn} &=& K_{mn}K_{ij}\,, \quad {\rm if\ all\ indices}\ i,j,m,n\
{\rm different}\,,\\
K_{ij}a^{(\dagger)}_j&=&a^{(\dagger)}_i K_{ij}.
\end{eqnarray}
Here
$\nu$ is a constant related to the Calogero coupling constant, while
$K_{ij}$  are the elementary permutation
operators of the $S_N$ exchange algebra. We use the standard convention
that square brackets denote commutators and curly brackets anticommutators.

To make contact between the $SH_N(\nu)$ algebra and the
Calogero model\footnote{
For a review of the Calogero model as a classical and quantum integrable
model, see \cite{Ol1}.}
 one has to use the following ``Calogero
realization'' of $SH_N(\nu)$:

\begin{equation}
a_i = \frac{1}{\sqrt{2}} (x_i + D_i)\,, \quad
a^\dagger_i = \frac{1}{\sqrt{2}} (x_i - D_i)\,,
\end{equation}
with

\begin{equation}
D_i =\frac{\partial }{\partial x_i}+\nu \sum_{j\neq i}(x_i -x_j)^{-1}
(1-K_{ij})\,.
\end{equation}
The real Calogero coordinates $x_i$ and the so-called Dunkl derivatives $D_i$
\cite{Du1} can be shown, by a direct calculation, to satisfy the commutation
relations \cite{Po1,Br1}:

\begin{equation}
[x_i, x_j] = [D_i, D_j] = 0, \hskip .3truecm [D_i, x_j] = A_{ij}.
\end{equation}

The crucial observation that establishes the relation with the
Calogero model is that the Calogero Hamiltonian $H_{Cal}$ is related
to the operator

\begin{equation}
\label{eq:ham}
H = {1\over 2} \sum_{i=1}^N \{a_i \,,a^\dagger_i\} \,.
\end{equation}
This operator, like the ordinary harmonic oscillator Hamiltonian,
obeys the standard relations

\begin{equation}
[H, a^\dagger_i\,] = a^\dagger_i\,,\qquad [H, a_i\,] = -a_i \,\,.
\end{equation}
The operator $H$ is called the universal Calogero Hamiltonian because
it also can be used to describe spinning Calogero models \cite{Mi1,Do}.
The precise relationship with the original Calogero Hamiltonian $H_{Cal}$
involves a simple similarity transformation
followed by a restriction to the subspace of totally symmetric
wave functions. This construction
allows one to construct all eigen wave functions of the model
as the Fock vectors

\begin{equation}
(a_1^\dagger)^{n_1}\ldots
(a_N^\dagger)^{n_N}|0\rangle \,.
\end{equation}
For more details, we refer to \cite{Br1}.

Our goal in this section is to investigate whether the $SH_N(\nu)$
algebra defined in
eqs.~(1)-(5) can lead to new realizations of the Virasoro algebra.
Our starting point is the following Ansatz for the Virasoro
generators
:

\begin{equation}
\label{eq:Ansatz}
L_{-n} = \sum_{i=1}^N \left ( \alpha (a_i^\dagger)^{n+1} a_i +
\beta a_i (a_i^\dagger)^{n+1}
+(\lambda-{1\over 2}) (n+1) (a_i^\dagger)^{n} \right )\,,
\end{equation}
where $\alpha, \beta$ and $\lambda$ are arbitrary
parameters\footnote{For
$N=1$ and $\alpha=\beta=1/2$
the parameter $\lambda$ coincides with the $\lambda$ of \cite{Be1}.}.
Note that for a vanishing value of the parameter $\nu$
we have that $[a_i,a_j^\dagger]=\delta_{ij}$ and
the above generators correspond to a direct sum of the
$N$ standard vector field generators of the Virasoro algebra.
Remarkably, it turns out that the Ansatz (\ref{eq:Ansatz})
also works for non-vanishing values of $\nu$. In order to
prove this it is convenient to sum over all the modes and
to rewrite the Ansatz for the Virasoro generators in terms
of a parameter function $\xi(a_i^\dagger)$ as

\begin{equation}
\label{Le}
L_\xi = \sum_{i=1}^N \left ( \alpha\ \xi(a_i^\dagger) a_i + \beta\
a_i\xi(a_i^\dagger)
+ (\lambda - {1\over 2} )
{\partial\over \partial a_i^\dagger} \xi (a_i^\dagger)
\right )\,.
\end{equation}

The proof that the Ansatz (\ref{Le}) satisfies the Virasoro algebra
commutation relations requires the following three identities

\begin{eqnarray}
\label{eq:id3}
\sum_{i=1}^N \xi_1(a_i^\dagger) [ a_i, \xi_2(a_j^\dagger)] - (
1\leftrightarrow 2)
&=& \xi_1(a_j^\dagger){\partial\over \partial a_j^\dagger}\xi_2(a_j^\dagger)
- (1 \leftrightarrow 2)\,,\\
\label{eq:id5}
\sum_{i=1}^N  [ a_i, \xi_1(a_j^\dagger)]\xi_2(a_i^\dagger) - (
1\leftrightarrow 2)
&=& {\partial\over \partial a_j^\dagger}
\xi_1(a_j^\dagger)\,\xi_2(a_j^\dagger) - (1\leftrightarrow 2)\,,\\
\label{eq:id4}
\sum_{i,j=1}^N \left [ [a_i, \xi_1 (a_i^\dagger) ], [a_j, \xi_2 (a_j^\dagger) ]
\right ] &=& 0\,,
\end{eqnarray}
where $\xi_1$ and $\xi_2$ are arbitrary Laurent series.
The proof of the identities (\ref{eq:id3}), (\ref{eq:id5}) and
(\ref{eq:id4}) is given in
appendix A. It is based on the following useful formula:
\begin{equation}
\label{eq:com}
[a_i , f(a^\dagger )] = \frac{\partial}{\partial a^\dagger_i }f(a^\dagger ) -
\nu \sum_{l=1}^N (a_i^\dagger -a_l^\dagger )^{-1} [K_{il} ,f(a^\dagger ) ]\,.
\end{equation}
This formula is a direct consequence of the basic commutation
relations (1) and can be
easily proven by expanding $f(a^\dagger )$ in a
Laurent series in $a^\dagger_i$.
Alternatively, one can observe that the right-hand-side of (\ref{eq:com})
$(i)$ respects the Leibniz rule, $(ii)$ vanishes when $f=const$ and
$(iii)$ reduces to the basic commutation relation (1) for $f=a^\dagger_i$.
Combined altogether the properties $(i)$-$(iii)$ prove that the formula
(\ref{eq:com}) is valid for an arbitrary polynomial in $a^\dagger_i$
and $(a^\dagger_i )^{-1}$.

Note that the right-hand-side  of (\ref{eq:com}) remains regular for regular
functions $f(a^\dagger )$, i.e. the poles in $(a^\dagger_i -a^\dagger_l )$
cancel due to the commutator with $K_{il}$.
It is worth mentioning that the formula (\ref{eq:com}) can be used for a
simple derivation of the Dunkl derivative and its further generalizations
given in \cite{Br2} as the action of the $a_i$ - type operators in the Fock
modules with the vacuum states satisfying the conditions $a_i |0\rangle =0$,
$K_{ij} |0\rangle = T_{ij} |0\rangle $ where the matrices $T_{ij}$
realize some representation $t$ of the symmetric group $S_N$ which acts
on the
vacuum vector(s) $|0\rangle$. The Dunkl derivatives then correspond to the
trivial representation of $S_N$, $T_{ij}=1$, while the derivatives introduced
in \cite{Br2} correspond to a general representation $t$.

We now proceed with the proof that the Ansatz (\ref{Le})
satisfies the Virasoro algebra commutation relations. For simplicity,
we first consider the special case with $\lambda = 1/2$.
Using the standard Leibniz rule we find that

\begin{eqnarray}
[L_{\xi_1}, L_{\xi_2}] &=& \sum_{i,j=1}^N \biggl (
\alpha^2 \xi_1(a_i^\dagger) [a_i, \xi_2(a_j^\dagger)] a_j -
\beta^2 a_j [a_i,\xi_1(a_j^\dagger)]\xi_2(a_i^\dagger)\nonumber\\
&&+ \alpha\beta [\xi_1(a_i^\dagger)a_i, a_j\xi_2(a_j^\dagger)] \biggr )
- (1 \leftrightarrow 2)\,.
\end{eqnarray}
The third term at the right-hand-side can be rewritten as
\begin{eqnarray}
\alpha\beta\sum_{i,j=1}^N \biggl (&&
{1\over 2} [\xi_1(a_i^\dagger)a_i, \xi_2(a_j^\dagger)a_j]
+ {1\over 2} [ a_j\xi_1(a_j^\dagger), a_i\xi_2(a_i^\dagger)]\nonumber\\
&& - \biggl [ [a_i, \xi_1(a_i^\dagger)], [a_j, \xi_2(a_j^\dagger)]
\biggr ]\biggr) - (1\leftrightarrow 2)\,.
\end{eqnarray}
The last term in this expression
vanishes due to the identity (\ref{eq:id4}) while the first two terms
are identical to the contribution from the $\alpha^2$ and $\beta^2$ terms.
Application of the identities (\ref{eq:id3}),
(\ref{eq:id5}) to the remaining terms gives the result

\begin{equation}
[L_{\xi_1}, L_{\xi_2}] =
(\alpha + \beta)\sum_{i=1}^N\left ( \alpha \xi_{1,2}(a_i^\dagger)a_i +
\beta a_i \xi_{1,2}(a_i^\dagger)\right )
\end{equation}
with the parameter $\xi_{1,2}(a_i^\dagger)$ given by

\begin{equation}
\xi_{1,2}(a_i^\dagger) =
\xi_1(a_i^\dagger) {\partial\over\partial a_i^\dagger}
\xi_2(a_i^\dagger)
-\xi_2(a_i^\dagger) {\partial\over\partial a_i^\dagger}
\xi_1(a_i^\dagger)\,.
\end{equation}
We conclude that the Ansatz (\ref{eq:Ansatz}) leads to the Virasoro algebra
provided that
\begin{equation}
\alpha + \beta = 1\,.
\end{equation}
This result can easily be extended
to other values of $\lambda$ with $\lambda \ne 1/2$
by the use of only the identities (\ref{eq:id3}) and (\ref{eq:id5}).
\bigskip

Thus it is shown that there exists a two-parametric
class\footnote{Particular examples of this class
has been known to other authors: S.~Isakov and J.~Leinaas
have studied the case
with $\alpha=\beta=1/2$, while the fact that the Virasoro algebra
closes for the
cases with $\alpha=0$ or $\beta=0$ was pointed out to us by
A.~Polychronakos (private communications).}
of realizations of the
Virasoro algebra constructed from the Calogero oscillators which are
the generating elements of $SH_N(\nu)$.
We now make some comments to the above result:
\bigskip

{\bf (1)}\ The proof that the generators (\ref{Le}) form the Virasoro
algebra is based entirely on the commutation relations (1)-(5)
and is independent of any particular realization of this algebra.
By their construction the generators are invariant under the action
of the symmetric group

\begin{equation}
\label{Sym}
K_{ij} L_\xi = L_\xi K_{ij}\,.
\end{equation}
Among other things this means that the Virasoro algebra closes and
the above
$S_N$-invariance property remains
valid if the parameters $\alpha$ and $\lambda$ are not just pure
numbers but depend on any combination of group algebra elements
of $S_N$ which are
$S_N$ invariant themselves, i.e.
\begin{equation}
 \alpha =
\alpha (T_n )\,;\qquad
\lambda =
\lambda (T_n )\,,
\end{equation}
\begin{equation}
T_n =
 \sum_{i_1\neq i_2\neq \ldots\neq i_n }K_{i_1 i_2}
K_{i_2 i_3}\ldots K_{i_n i_1}\,.
\end{equation}
In the discussion below we will only need the case that $\alpha$ is constant
and $\lambda$ is at most linear in $K_{ij}$:
\begin{equation}
 \alpha =
\alpha_0\,,\qquad
\lambda =
\lambda_0 +\lambda_1 \ \sum_{i\ne j} K_{ij} \,,
\end{equation}
where $\alpha_0$, $\lambda_0$ and $\lambda_1$ are some constants.
This freedom allows us to define the $L_0$
generator of the Virasoro algebra
in such a way that it satisfies the properties of a $Z$-grading operator
for
the whole enveloping algebra of $SH_N(\nu)$, i.e.~it satisfies the
same identities as the universal Calogero Hamiltonian (see (10)).

The generators $\{L_1, L_0, L_{-1}\}$
which are given by

\begin{eqnarray}
\label{eq:bos}
L_1 &=& \sum_{i=1}^N a_i\,, \nonumber\\
L_0 &=& \sum_{i=1}^N \biggl ( \alpha a_i^\dagger a_i + (1-\alpha)
a_ia_i^\dagger\biggr )
+ (\lambda -{1\over 2})N\,,\\
L_{-1} &=& \sum_{i=1}^N \biggl ( \alpha (a_i^\dagger)^2 a_i
+ (1-\alpha) a_i (a_i^\dagger)^2 \biggr )
+ 2(\lambda - {1\over 2}) \sum_{i=1}^N a_i^\dagger\,,\nonumber
\end{eqnarray}
form a $sl(2)$ subalgebra of the Virasoro algebra.
$L_1$ coincides with the annihilation operator
of the center of mass degree of freedom while
$L_0$ is nothing else than the universal Calogero Hamiltonian (\ref{eq:ham})
modulo terms which become a
constant when acting on the subspace of symmetric wavefunctions, i.e.:

\begin{equation}
L_0 - H = (\lambda - \alpha )N - \nu(\alpha - {1\over 2})
\sum_{i\ne j} K_{ij}\,.
\end{equation}
One can fix the free parameteres $\alpha$ and $\lambda$ to be
\begin{equation}
\label{adjust}
\alpha = \alpha_0 \qquad \lambda=\lambda_0 +\nu N^{-1} (\alpha
-{1\over 2} ) \sum_{i\ne j} K_{ij}\,,  \end{equation} so that $L_0 - H$
becomes a pure constant \begin{equation} L_0 - H =
(\lambda_0 - \alpha_0 )N\,.
\end{equation}

We denote the $sl(2)$ subalgebra spanned by
$\{L_{-1},L_0,L_1\}$ as the ``horizontal'' $sl(2)$ algebra to distinguish
it from the ``vertical'' $sl(2)$ algebra of \cite{Pe1} which also acts
on the states of the Calogero model and is spanned by the generators

\begin{equation}
\label{eq:vert}
B^+_2 ={1\over 2}\sum_{i=1}^N (a_i^\dagger)^2,\hskip .3truecm
 B_2^0 = H, \hskip .3truecm
B^-_2 ={1\over 2}\sum_{i=1}^N (a_i)^2\,.
\end{equation}
Remarkably, the (shifted by a constant) Calogero Hamiltonian serves as the
Cartan subalgebra generator of both $sl(2)$ algebras.  We have not been able
to extend the vertical $sl(2)$ algebra to a Virasoro algebra.

The excited states of the Calogero model can be classified
both with respect to ``vertical''  and ``horizontal'' $sl(2)$
algebras. Our results imply that for the latter case these representations
extend to appropriate representations of the whole reductive
part of the Virasoro algebra spanned by the
$L_n$ with $n\le 1$. It is
worth mentioning that both the ``vertical'' $sl(2)$ algebra and the
``horizontal'' Virasoro algebra are generated by symmetric
combinations of the
Calogero creation and annihilation operators $a_i$ and $a_i^\dagger$
(cf.~(\ref{Sym})) so that the Virasoro algebra under consideration does
indeed leave
invariant the subspace of totally symmetric wavefunctions of the Calogero
model.
\bigskip

{\bf (2)}\ So far we have used the Calogero
realization where the creation and annihilation operators
are expressed in terms of the real coordinates $x_i$
underlying the Calogero model. However, one can use other
representations of the same algebra equally well. For example,
the Virasoro commutation relations remain valid for all other
representations found in \cite{Br2}. A particular useful realization is
that which for $\nu =0$ reduces to the standard holomorphic
representation and thus can be expected to be relevant to conformal
field theory. To be specific, consider the ``holomorphic realization'' of the
$SH_N(\nu)$ algebra given by:

\begin{equation}
a_i = D^z{}_i\,, \hskip 1.5truecm a_i^\dagger = z_i\,,
\end{equation}
where the $z_i$ are $N$ complex coordinates and $D^z{}_i$ is the
complex Dunkl derivative:

\begin{equation}
\label{eq:Dunkl}
D^z{}_i =\frac{\partial }{\partial z_i}+\nu \sum_{j\neq i}(z_i -z_j)^{-1}
(1-K_{ij})\,.
\end{equation}

In the holomorphic representation the Virasoro generators take the
following form:

\begin{equation}
\label{eq:Lhol}
L_\xi = \sum_{i=1}^N \left ( \alpha \xi (z_i) D^z{}_i +
(1-\alpha) D^z{}_i \xi(z_i) +
(\lambda - {1\over 2}) {\partial\over \partial z_i} \xi (z_i)
\right )\,.
\end{equation}
Inserting (\ref{eq:Dunkl}) into (\ref{eq:Lhol}), one
can write $L_\xi$ as follows:

\begin{eqnarray}
\label{LKZ}
L_\xi = &&\sum_{i=1}^N \left ( \alpha \xi (z_i) \partial^{KZ}_i +
(1-\alpha) \partial^{KZ}_i \xi(z_i)  +
(\lambda - {1\over 2}) {\partial\over \partial z_i} \xi (z_i)
\right )\nonumber\\
&&+ \nu(1-2\alpha)\sum_{i\ne j } \xi(z_i) {1\over z_i - z_j} K_{ij}\,,
\end{eqnarray}
where the Knizhnik-Zamolodchikov-type derivatives $\partial_i^{KZ}$
are defined by

\begin{equation}
\partial_i^{KZ} =
\frac{\partial }{\partial z_i}+\nu \sum_{j\neq i}(z_i -z_j)^{-1}\,.
\end{equation}

We observe that for $\alpha = 1/2$ all $K$-dependent terms vanish.
Since the Knizhnik-Zamolodchikov derivatives satisfy the ordinary
Heisenberg algebra $[\partial_i^{KZ}, z_j]=\delta_{ij}$ one is left, for
$\alpha = 1/2$, with
the standard realization of the Virasoro algebra\footnote{One could try
to use this observation to give an alternative proof of the existence of the
Virasoro algebra with $\alpha ={1\over 2}$ for other realizations of the
$SH_N(\nu)$ algebra.  Indeed, if there exists an operator $U$ intertwining the
holomorphic and some other representation of the $SH_N(\nu)$ algebra (e.g.~the
Calogero representation we used above) then the proof will become trivial for
this other representation as well.  Unfortunately, we do not know whether
there exists such an operator intertwining between the holomorphic and the
Calogero representation.  Note that, since $L_0$ is the universal Calogero
Hamiltonian the knowledge of such an operator $U$ would imply in particular
an explicit solution of the Calogero model via reduction to an ordinary
harmonic oscillator problem.}. It is worth mentioning that the fact that
the $K$-dependence trivializes for $\alpha = 1/2$
in the holomorphic representation
for the Calogero Hamiltonian was already observed in \cite{Br3} when discussing
the interplay between the Calogero model and anyons.
\bigskip

{\bf (3)}\
There exists the following important difference between the horizontal
$sl(2)$ algebra (\ref{eq:bos}) and the vertical $sl(2)$ algebra
(\ref{eq:vert}) with respect to the dependence of the generators
on the center of mass and relative coordinates.
In the vertical $sl(2)$ the center of
mass degrees of freedom decouple from the relative motion degrees of
freedom for arbitrary $N$ in view of the following relation

\begin{equation}
\sum_{i=1}^N \{X_i,Y_i\} = {1\over N}\left (
\sum_{i<j}^N \{X_i-X_j,Y_i-Y_j\} + \{\sum_{i=1}^N X_i,
\sum_{j=1}^N Y_j\}\right )\,.
\end{equation}
On the contrary, it turns out that for the horizontal $sl(2)$ algebra in
the holomorphic representation we have a nontrivial
mixture of the centre of mass degrees of freedom and the
relative motion ones.

The center of mass coordinate $y$ and
the relative coordinates $\tilde z_i$ are defined by

\begin{eqnarray}
y &=& {1\over N}\sum_{i=1}^N z_i\,, \quad \tilde z_i = z_i - y\,,\\
{\partial\over\partial y} &=& \sum_{i=1}^N {\partial\over \partial z_i}\,,
\quad {\partial\over \partial \tilde z_i} = {\partial\over \partial z_i}
- {1\over N} {\partial\over \partial y}\,.
\end{eqnarray}
Note that $\sum_{i=1}^N {\tilde z_i} = \sum_{i=1}^N \partial/\partial
{\tilde z_i} = 0$ and that $\partial/\partial
{\tilde z_i}{\tilde z_j} = \delta_{ij}-1/N$.
The expressions
for the generators of the horizontal $sl(2)$ algebra in terms of $y$ and
$\tilde z_i$, when the arbitrary parameters are fixed according to
(\ref{adjust}), are given by
\begin{eqnarray}
\label{eq:mixing}
L_1 &=& {\partial\over \partial y}\,,\nonumber\\
L_0 &=& y{\partial\over \partial y}
+ \sum_{i=1}^N \tilde z_i {\partial\over \partial \tilde z_i}
+ N\left (\lambda_0 + {1\over 2} - \alpha +{1\over 2}\nu
(N-1) \right )\,,\nonumber\\
L_{-1} &=& y^2{\partial\over\partial y}
+ 2\sum_{i=1}^N y\tilde z_i {\partial\over \partial \tilde z_i}
+ {1\over N}\sum_{i=1}^N \tilde z_i^2 {\partial\over\partial y}
+ \sum_{i=1}^N \tilde z_i^2 {\partial\over\partial \tilde z_i}
\nonumber\\
&+& \nu (1-2\alpha )\sum_{i\neq j} \tilde z_i K_{ij}+
2Ny \left (\lambda_0 + {1\over 2} - \alpha + {1\over 2}\nu
(N-1)\right )\,.
\end{eqnarray}

We observe that the relative coordinate
operators $\tilde z_i, {\partial/\partial\tilde z_i}$
all occur in the $z$ scale-invariant combination
$\tilde z_i\partial/\partial\tilde z_i$  except for three terms
in $L_{-1}$.
One of these three terms leads to a nontrivial mixing of the center of
mass degrees of freedom and the relative ones in the Calogero model.
To get rid of this term, one can consider the
limiting procedure where $y \rightarrow y$
and $\tilde z_i\rightarrow \delta\tilde z_i$ with $\delta \rightarrow 0$
which does not affect the commutation relations of the
$sl(2)$
algebra as can be checked
easily by using (\ref{eq:mixing}). We observe that after this
limiting procedure the relative coordinate-dependent operators
always occur in the combination

\begin{equation}
\label{bl}
\bar \lambda =\sum_{i=1}^N \tilde z_i {\partial\over \partial \tilde z_i}
+ N (\lambda_0  +{1\over 2} - \alpha  + {1\over 2}\nu (N-1))\,.
\end{equation}
For this particular degenerate
realization the relative coordinate operators thus behave as
inner coordinates which only effect the conformal weight.

The above limiting procedure can be applied to the whole Virasoro
algebra as well. One may verify that after taking the limit
$\delta \rightarrow 0$ the generators following from (\ref{LKZ})
all take the standard form
\begin{equation}
\label{lim}
L_{-n}=y^{n+1}{\partial\over \partial y} +\bar \lambda (n+1)y^n\,,
\end{equation}
with the conformal weight $\bar{\lambda}$ given by (\ref{bl}).

Finally, we note that the center of mass coordinate $y$ plays an
important role in defining, in a consistent way, the Virasoro
generators $L_n\ (n>0)$ which involve negative powers of $z_i$.
Specifically, to define inverse powers of $z_i$ one first uses
the decomposition $z_i = y + \tilde z_i$ and then expands all
expressions in powers of the relative coordinates $\tilde z_i$,
e.g.~$z_i^{-1} = y^{-1}\sum_{n=0}^\infty (-)^n  ({\tilde z_i\over y})^n$.
In particular, it is convenient to use this approach in deriving the
limiting form (\ref{lim}). The same approach is used in the analysis
of the $W_\infty$-type generalizations which are discussed in section 4.

\vspace{.5cm}
\section{Super Extension}
%\noindent{\bf 3. Super Extension }

\bigskip

It is natural to extend the results obtained in the previous section to the
supersymmetric case, thereby extending the Virasoro algebra to a
super-Virasoro algebra. Supersymmetric extensions of the Calogero model were
recently discussed in \cite{Fr1,Br3}. In the following we will frequently
make use of the results of \cite{Br3}.

Our starting point is the
supersymmetric extension of the $SH_N(\nu)$ algebra which is given by the
direct product of $SH_N(\nu)$ with the Clifford algebra $C_{2N}$ with
generating elements $\theta_i$ and $\theta_i^\dagger$:

\begin{equation}
\{\theta_i, \theta_j^\dagger\} = \delta_{ij}\,.
\end{equation}
The operator $K_{ij}$ is assumed to commute with $\theta_i$ and $
\theta_i^\dagger$. Note that the fermionic permutation operators
$K_{ij}^\theta$ can be realized as\footnote{
We observe that for $N=2$ the operator $K_{12}^\theta$ equals the Klein
operator $K$ which occurs in the discussion of the super-$W_\infty(\lambda)$
algebra \cite{Be1}. Indeed, we can write $K = 1 - 2 \theta \theta^\dagger$
with $\theta^{(\dagger)} \equiv {1\over \sqrt 2}(\theta_1^{(\dagger)} -
\theta_
2^{(\dagger)})$.
The Klein operator $K$ satisfies $K^2 = 1$ and anticommutes with
$\theta^{(\dagger)}$.}
:

\begin{equation}
K_{ij}^\theta = 1 -(\theta_i - \theta_j)(\theta_i^\dagger - \theta_j^\dagger)
\,.
\end{equation}
These operators commute with $a_i, a_i^\dagger$ and have the following
standard permutation relations with $\theta_i,\theta_i^\dagger$:

\begin{equation}
\theta_i^{(\dagger)}K_{ij}^{\theta} =
K_{ij}^{\theta} \theta_j^{(\dagger)}\,, \quad (K_{ij}^\theta)^2 = 1\,.
\end{equation}
In addition, the operators $K_{ij}^\theta$ satisfy the properties
(2)-(4). One can also define the total permutation operators

\begin{equation}
K_{ij}^{\rm tot} = K_{ij}K_{ij}^\theta\,,
\end{equation}
which exchange both bosonic and fermionic coordinates
simultaneously.

In \cite{Fr1,Br3} the explicitly supersymmetric form of the super-Calogero
Hamiltonian $H_s$ was given. In particular, the construction of
\cite{Br3} was based upon an $osp(1,2)$ supersymmetric extension
of the vertical $sl(2)$ algebra\footnote{We have reintroduced the
frequency parameter $\omega_F$ of \cite{Br3} and taken it to be
equal to $\omega_F = -{1\over 2}$.}

\begin{eqnarray}
\label{eq:Hs}
H_s &=& \{Q, Q^\dagger\}\,,\nonumber\\
&=& {1\over 2} \sum_{i=1}^N  \{a_i, a_i^\dagger\} +
{1\over 4}\sum_{i=1}^N
[\theta_i, \theta_i^\dagger] + {1\over 2}\nu \sum_{i\ne j}
K_{ij}^{\rm tot}\,,
\end{eqnarray}
where

\begin{equation}
Q= \sum_{i=1}^N \theta_i^\dagger a_i\,, \quad Q^\dagger =
\sum_{i=1}^N \theta_i a_i^\dagger\,.
\end{equation}
The generators $Q, Q^\dagger$ can be interpreted as odd generators
of the $osp(1,2)$ superextension of the vertical $sl(2)$ algebra.
Note that the last term in (\ref{eq:Hs}), when acting on
the subspace of totally symmetric wavefunctions, reduces to a constant.
One may verify that the hamiltonian
\begin{equation}
H=
{1\over 2} \sum_{i=1}^N  \{a_i, a_i^\dagger\} +
{1\over 4}\sum_{i=1}^N
[\theta_i, \theta_i^\dagger]
\end{equation}
satisfies the
commutation relations:

\begin{eqnarray}
\label{eq:comm}
[H, a_i^\dagger] &=& a_i^\dagger\,, \hskip 1truecm [H, a_i] = -a_i
\,,\\
{[} H, \theta_i^\dagger {]} &=&
-{1\over 2} \theta_i^\dagger\,, \hskip
.4truecm
[H, \theta_i ] =  {1\over 2} \theta_i\,.
\end{eqnarray}

In order to describe our Ansatz for the super-Virasoro generators, it is
convenient to introduce
the  ``super-Dunkl derivative'' ${\cal D}_i$. In \cite{Br3} it was shown
that there exists a one-parametric class of superderivatives which all fulfill
the basic relations
\begin{eqnarray}
\label{eq:commd}
\{ {\cal D}_i, {\cal
D}_j \} &=& 2\delta_{ij} D^\theta_i\,,\\ \{ {\cal D}_i, \theta_j \} &=&
\delta_{ij}\,,
\end{eqnarray}
where $D^\theta_i $ is some supersymmetric
extension of the ordinary (bosonic) Dunkl derivative. All these derivatives
were shown in \cite{Br3} to be related to each other by a similarity
transform. Remarkably, it turns out that only one
particular member of this class leads
to the desired superextension of the
Virasoro algebra\footnote{
Note that the similarity transform of \cite{Br3} does not
leave invariant the Ansatz we use below.}. This
particular member is exactly the one which is covariant with respect to
the global supersymmetry rules

\begin{equation}
\delta z_i = \theta_i\epsilon\,, \quad \delta \theta_i = \epsilon\,.
\end{equation}

To be precise, for the realization of the
super-Virasoro algebra, we need the following special derivative\footnote{
This derivative equals the super-Dunkl derivative ${}^\alpha{\cal D}_i$ of
\cite{Br3} with $\alpha = \nu$.}:
\begin{equation}
{\cal D}_i =
{\cal D}^0_i
 +\nu\sum_{j\neq i} {\theta_{ij}\over z_{ij}}(1-K_{ij}^{{\rm tot}})\,.
\end{equation}
where
\begin{equation}
{\cal D}^0_i = {\partial\over \partial \theta_i} +
\theta_i {\partial\over \partial z_i}
\end{equation}
is the ordinary supercovariant derivative and where we have introduced
the supersymmetric line elements

\begin{equation}
z_{ij} =  z_i - z_j - \theta_i\theta_j\,, \quad
\theta_{ij} = \theta_i - \theta_j\,.
\end{equation}
The supercovariant derivatives of these line-elements are given by

\begin{equation}
{\cal D}_i^0 z_{ij} = {\cal D}_j^0 z_{ij} =  \theta_{ij}
\,, \quad
{\cal D}_i^0 \theta_{ij} = - {\cal D}_j^0 \theta_{ij} = 1\,.
\end{equation}
A noteworthy property of the super-Dunkl derivative ${\cal D}_i$
is that because of the
factor of $\theta_i -\theta_j $ in front of $K^{{\rm tot}}_{ij}$ one can
equally well use both $K_{ij} $ and $K_{ij}^{{\rm tot}}$ in its
definition. The expression for the bosonic derivative $D_i^\theta$
follows from the anticommutator of two super-Dunkl derivatives and
is given by:

\begin{equation}
\label{bsd}
D_i^\theta %&=&
={\partial\over \partial z_i} + \nu\sum_{l\ne i}\biggl [
{1\over z_{il}}(1-K_{il}^{{\rm tot}}) - {\theta_{il}\over z_{il}}
K_{il}^{{\rm tot}}\left({\cal D}_i - {\cal D}_l -\nu\sum_{j\ne i,j\ne l}
{\theta_{lj}\over z_{lj}}K_{lj}^{{\rm tot}}
\right)\biggr ]\,.
%\nonumber\\
%&&+ \nu^2 \sum_{k,l\ne i} {\theta_{ik}\theta_{kl}\over z_{kl}z_{li}}
%(1-K_{il}^{{\rm tot}})K_{ik}^{{\rm tot}}
\end{equation}

One easily finds that
\begin{equation}
\label{al}
\left [  {\cal D}_i, z_j \right ] =\delta_{ij}\biggl (
\theta_i + \nu \sum_{k\ne i =1}^N \theta_{ik}K_{ik}\biggr )
 -\nu \theta_{ij} K_{ij} \,.
\end{equation}
This formula is a special case of a more general formula which gives the
commutator of ${\cal D}_i$ with an arbitrary function
$f(z,\theta )$ of a single
superargument $(z,\theta )$. Using the convention $f_i \equiv
f(z_i ,\theta_i )$ and the abbreviation
\begin{equation}
y_{ij} \equiv {\theta_{ij}\over z_{ij}}\,,
\end{equation}
the general formula is given by
\begin{equation}
\label{1}
\left [  {\cal D}_i , f _i \right ] ={\cal D}^0(f)_i
+\nu\sum_{i\neq j}(f_i -f_j )\,y_{ij}K_{ij}^{{\rm tot}}
\,,
\end{equation}

\begin{equation}
\label{2}
 \left [  {\cal D}_i, f _j \right ] =
-\nu(f_i -f_j )\,y_{ij}K_{ij}^{{\rm tot}}
\,,\qquad i\ne j
\end{equation}
\begin{equation}
\label{tr}
\sum_i \left [  {\cal D}_i, f _i \right ] =
\sum_i {\cal D}^0(f)_i \,.
\end{equation}

For the convenience of the reader we give below some useful identities
that are obeyed by the quantities $y_{ij}$:

\begin{eqnarray}
\label{eq:y-id}
y_{ij}&=&y_{ji}\,,\qquad y_{ij} y_{kl}=-y_{kl}y_{ij}\,,\nonumber\\
y_{il}y_{lj}&+&y_{jl}y_{ij} +y_{ij}y_{il}=0\,,\\
{\cal D}^0_i y_{ij}&=& -{\cal D}^0_j y_{ij}={1\over z_{ij}}\,.\nonumber
\end{eqnarray}

We now proceed with our Ansatz
for the
(super)generators of the $(N=2)$ super Virasoro algebra which
is given by

\begin{equation}
\label{svir}
L_\xi =\sum_i ( {\cal D}_i\xi_i {\cal D}_i -\frac{1}{2}{\cal D}^0
(\xi )_i {\cal D}_i +\frac{1}{2} \lambda \xi_i^\prime )\,,
 \end{equation} \begin{equation} Q_\epsilon
=\sum_i \epsilon_i {\cal D}_i +\lambda \sum_i {\cal D}^0 (\epsilon )_i\,,
\end{equation}
where $\xi (z,\theta )$ and $\epsilon (z,\theta )$ are arbitrary
commuting parameters and $f^\prime (z,\theta ) \equiv $
$ {\partial \over\partial z }f(z,\theta )$.

One can check that the following (anti-)commutation relations are true
\begin{equation}
\label{QQ}
\{Q_{\epsilon_1 } ,Q_{\epsilon_2 } \} = L_{\xi_{1,2}}\\,
\end{equation}
\begin{equation}
\label{LQ}
[L_{\xi} , Q_{\epsilon }]  = Q_{\bar \epsilon}\\,
\end{equation}
where
\begin{equation}
\xi_{1,2} =2\epsilon_1 \epsilon_2
\end{equation}
and
\begin{equation}
\label{ps}
\bar\epsilon =  \xi \epsilon^\prime
-\frac{1}{2} \epsilon \xi^\prime
+\frac{1}{2} {\cal D}^0 (\xi ) {\cal D}^0 (\epsilon)\,.
\end{equation}

The proof of the anti-commutation relation
(\ref{QQ}) is relatively simple. To illustrate how
it works we consider the case with $\lambda =0$. A straightforward
calculation gives
\begin{equation}
Q_\epsilon Q_\epsilon =
\sum_{i,j}
\epsilon_i \D_i
\epsilon_j \D_j =
\sum_{i,j}
(\epsilon_i
\epsilon_j \D_i\D_j +
\epsilon_i [\D_i ,
\epsilon_j ]\D_j )\,.
\end{equation}
Taking into account the fact that the parameter $\epsilon$ is commuting,
the basic anti-commutation relation (\ref{eq:commd})
of the super-Dunkl derivatives
and the formulae (\ref{1}), (\ref{2}) one
obtains after some simple algebra
\begin{equation}
\label{ac1}
Q_\epsilon Q_\epsilon =
\sum_{i}(
 \epsilon_i^2
\D_i^2 +
\epsilon_i \D^0_i(\epsilon_i )\D_i ) +
\nu \sum_{i\ne j}
\epsilon_i^2 y_{ij} K^{\rm tot}_{ij} ( {\cal D}_i - {\cal D}_j)\,.
\end{equation}
Analogously one finds
\begin{equation}
\label{ac2}
\sum_{i}
 \D_i \epsilon_i^2 \D_i =
\sum_{i}
(\epsilon_i^2
\D_i^2 +
 \D^0_i(\epsilon_i^2 )\D_i ) +
 \nu \sum_{i\ne j}
\epsilon_i^2 y_{ij} K^{\rm tot}_{ij} ({\cal D}_i - {\cal D}_j)\,.
\end{equation}
Combining (\ref{ac1}) with (\ref{ac2})
proves the relation (\ref{QQ}) for $\lambda =0$.
It is simple to generalize this relation to non-vanishing values of
$\lambda$.

The proof of the $[L_\xi,Q_\epsilon]$ commutator is much more
involved and is described in appendix B. Let us note that as
in the bosonic case
the fact that the generators $Q$ and $L$ form a closed
super Virasoro algebra is representation independent,
i.e.~it follows only from the basic commutation relations
(\ref{eq:commd}),(53) and (\ref{al}).
Note that
from the relations (\ref{QQ}) and (\ref{LQ}) it follows
that the $L$ generators
satisfy the commutation relations of the Virasoro algebra.

Like in the bosonic case $L_0$ can be identified with a $Z$-grading operator
by adjusting the $\lambda$ parameter as follows:
\begin{equation}
\lambda =\lambda_0 +\nu N^{-1} \sum_{i\neq j} (K_{ij}^{\rm tot} -1)\,.
\end{equation}
The generators $L_{1}$, $L_0$ and $L_{-1}$ of the $sl_2$ subalgebra
of Virasoro which correspond respectively to the parameters
$\xi =$ 1, $z$, and $z^2$ in (\ref{svir})
then take the form
\begin{equation}
L_1 =\sum_{i=1}^N {\partial\over \partial z_i}\,,
\end{equation}
\begin{equation}
L_0 =\sum_i z_i {\partial\over \partial z_i}+{1\over 2}
    \sum_i \theta_i {\partial\over \partial \theta_i}
+{1\over 2}\lambda_0 N\,,
\end{equation}
\begin{eqnarray}
L_{-1} &=&\sum_i z_i^2 {\partial\over \partial z_i}+
    \sum_i z_i \theta_i {\partial\over \partial \theta_i}
+\lambda_0 \sum_{i=1}^N z_i
-\nu \sum_{i\neq j}z_i K^{\rm tot}_{ij}\,
\nonumber\\
&+&\nu N^{-1} \sum_{i\neq j}K^{\rm tot}_{ij} \sum_{l=1}^N z_l
+\nu^2 \sum_{j\ne k \ne i}
z^2_i y_{ij}K_{ij}^{\rm tot} y_{jk}K_{jk}^{\rm tot}\,.
\end{eqnarray}
The $\nu^2$ contribution to $L_0$ from the
last term in the representation (\ref{LL}) for the Virasoro generators
vanishes due to the following identity
\begin{equation}
z_l y_{il}y_{lj} + z_j y_{jl}y_{ij} + z_i y_{ij}y_{il} = 0\,.
\end{equation}

The fermionic coordinates in $L_0$ only appear
as a shift of the conformal parameter $\lambda_0$ which involves the
fermionic number operator
$\sum_i \theta_i {\partial\over \partial \theta_i}$. This is not the case
for the $L_1$ generator which involves non-trivial mixings between
the relative and fermionic coordinates.

An interesting distinction between the bosonic and fermionic case
is that the supergenerators involve only one independent free
parameter, the conformal weight $\lambda$. Due to the identity (\ref{tr})
a term of the form $\sum_i {\cal D}_i\epsilon_i$ can be rewritten in
terms of $\sum_i\epsilon_i{\cal D}_i$ modulo a shift in the parameter
$\lambda$. Therefore, in the supersymmetric case, there is no room
for a second free parameter, like the parameter $\alpha$ in the
bosonic case.

When restricted to the subspace of $\theta$-independent functions, the
above generators of the $sl_2$ subalgebra of the super-Virasoro algebra
coincide with the bosonic generators (\ref{eq:bos}) for $\alpha =1$ (
and with the conformal weight parameters identified as
$\lambda$(bosonic) - 1/2 = 1/2 $\lambda$(fermionic) ) except for the
$\nu^2$-dependent term in $L_{-1}$.  This term contains an explicit
$\theta$-dependence.  Therefore, for $\nu \ne 0$, one cannot truncate
the $sl_2$ subalgebra of the super-Virasoro algebra to the sector of
operators acting on $\theta$-independent functions.

The super-Virasoro algebra under consideration is formulated here in
terms of $N=1$ superfield parameters analogous to the one-particle
formulation given in \cite{Be1}.  The algebra however is $N=2$
supersymmetric as can easily be seen from the component analysis: the
superfield generators $L_\xi$ and $Q_\epsilon$ involve one spin 2
current, {\sl two} spin 3/2 currents (one from $L_\xi$, and another one
from $Q_\epsilon$), and one spin 1 current (the $\theta$ - component in
$Q_\epsilon$).  In particular, the $u(1)$ component of the spin 1
current has the form
\begin{equation}
J_0 =    \sum_i  \theta_i {\partial\over \partial \theta_i}-\nu
\sum_{i\ne j}{\theta_{i}\theta_{j}\over z_{ij}}(1-K_{ij}^{{\rm tot}})\,.
\end{equation}

Along with the corresponding supergenerators the above spin 2 and spin
1 generators form the ``horizontal" $osp(2|2)$ subalgebra of the full
super Virasoro algebra. This algebra should be distinguished from the
``vertical" $osp(2|2)$ algebra which was considered
in \cite{Br3}.

\section{Conclusions}

In this paper we have shown that based on the Calogero model inspired
algebras $SH_N(\nu) $ and their superextensions one can construct
new realizations of the (super-) Virasoro algebra analogous to the standard
realization by (super) vector fields based on the ordinary
Heisenberg algebras (or, equivalently, the algebra of differential operators).
The algebra $SH_N(\nu) $ can be regarded as the associative algebra with
generating elements $a^\dagger_i$, $ a_j$ and $K_{ij}$. One can
consider the same algebra but now with the (super-) commutator
as the product law.
This leads to infinite-dimensional Lie (super)algebras which we
denote by $W_{N,\infty} (\nu)$ in analogy with the construction of ordinary
$W_\infty$-type algebras via commutators of elements of the enveloping algebra
of the Heisenberg algebra \cite{V,Winb,Winp}. Due to the results of this paper
these algebras contain the Virasoro subalgebra and therefore can be
regarded as an extended conformal algebra thereby justifying the name
$W_\infty$-type algebra. Note that the $W_{1+\infty}$ algebra can be
identified with the algebra $W_{1,\infty} (\nu)$ (the parameter $\nu$
drops out for $N=1$).

The  $W_{N,\infty} (\nu)\ (N\ge 2)$ algebras are rather large algebras
involving
an infinite number of higher spin generators of each spin.
An interesting question is whether there exists truncations of the
$W_{N,\infty} (\nu)$ algebra that still contain the Virasoro algebra.
There indeed exists such a truncated algebra.
This subalgebra is spanned by the symmetric elements
$w$ that commute with the permutation operators, i.e.
\begin{equation}
[K_{ij}\,,w ]=0\,.
\end{equation}
We denote this subalgebra of $W_{N,\infty} (\nu)$ by
$\tilde{W}_{N,\infty} (\nu)$. Since the Virasoro generators themselves
obey the symmetry conditions (\ref{Sym}), $\tilde{W}_{N,\infty} (\nu)$
contains the Virasoro algebra as a subalgebra.
The algebra
$\tilde{W}_{N,\infty} (\nu)$ contains infinitely
many higher-spin currents of each spin.

A natural question to ask is whether
$W_{1+\infty }$ is a proper subalgebra of $W_{N,\infty} (\nu)$.
The answer to this question seems to be negative. So far our
attempts
to embed $W_{1+\infty }$ into $W_{N,\infty} (\nu)$ as a
proper subalgebra were not successful.

The above results can be naturally extended
to the supersymmetric case. One can define the
superalgebras $SW_{N,\infty} (\nu)$ and $S\tilde{W}_{N,\infty} (\nu)$ based
on the superextensions considered in section 3. The
superalgebra $S\tilde{W}_{N,\infty} (\nu)$ is spanned by the
symmetric elements obeying the conditions
\begin{equation}
[K_{ij}^{{\rm tot}}\,,w ]=0\,.
\end{equation}

We believe that the generalized $W_\infty$-type algebras introduced
above may have interesting applications in the context of
conformal field theory, integrable systems and the quantum Hall effect.
We hope to discuss some of these applications in a future publication.
%\newpage
\vspace{1truecm}

\noindent {\bf Acknowledgments}
\vspace{.5truecm}

We thank L.~Brink, S.~Isakov, J.~Leinaas, and A.~Polychronakos for
useful discussions.
The work of E.B.~has been made possible by a fellowship of the Royal
Netherlands Academy of Arts and Sciences (KNAW). One of us (M.V.) would
like to thank Groningen University for hospitality.

%\appendix

%\vspace{1truecm}

%\noindent {\bf
\section*{Appendix A: Proof of the identities (\ref{eq:id3}),
(\ref{eq:id5})\\$\phantom{MMMMMM}$ and (\ref{eq:id4})}

\vspace{1truecm}

To prove the identities (\ref{eq:id3}), (\ref{eq:id5}) and
(\ref{eq:id4}) we need a particular case of (\ref{eq:com}) when
$f(a^\dagger )$ is a function of one argument $a^\dagger_i$
for some fixed $i$, i.e. $f(a^\dagger )=\phi (a^\dagger_i )$. For this
particular case one easily derives from (\ref{eq:com})
\begin{eqnarray}
\label{eq:com1}
[a_i , \phi(a^\dagger_j )]&=& \delta_{ij}\phi^\prime (a^\dagger_i )
+\nu\delta_{ij} \sum_{l=1}^N
(a_i^\dagger -a_l^\dagger
)^{-1}
(\phi (a^\dagger_i )-\phi (a^\dagger_l ) )K_{il}\nonumber \\
&-&\nu
(a_i^\dagger -a_j^\dagger )^{-1}
(\phi (a^\dagger_i ) -\phi (a^\dagger_j ) )K_{ij}\,,
\end{eqnarray}
where $\phi^\prime (x)=\frac{\partial}{\partial x}\phi (x)$.
Note that the commutation relations (1) are themselves a particular
case of (\ref{eq:com1}) for $\phi(x)=x$.

The proof of (\ref{eq:id3}) and (\ref{eq:id5})
is straightforward by observing that the
two $\nu$-dependent terms coming from the right-hand-side
of (\ref{eq:com1}) cancel against each other after the substitution of
(\ref{eq:com1}) into (\ref{eq:id3}), (\ref{eq:id5}).
The $\nu$-independent part of the resulting expression gives the
right-hand-side of (\ref{eq:id3}), (\ref{eq:id5}).

To prove the identity (\ref{eq:id4}) it is convenient to decompose
the left-hand-side of (\ref{eq:id4}),
\begin{equation}
X \equiv
\sum_{i,j=1}^N \biggl [ [a_i, \xi_1(a_i^\dagger ) ], [a_j,
\xi_2(a_j^\dagger ) ]\biggr ]
\end{equation}
into three parts
$X^{(0)}, X^{(1)}$ and $X^{(2)}$,
which are, respectively, independent of $\nu$,
linear in $\nu$ and quadratic in $\nu$. Due to (\ref{eq:com1})
we obviously have $X^{(0)}=0$. Using the shorthand notation
$\xi_{1i} = \xi_1(a_i^\dagger)$  and $\xi_{2i} = \xi_2(a_i^\dagger)$
we can write

\begin{eqnarray}
X^{(1)} &=&
\nu \sum_{i=1}^N [\xi_{1i}^\prime,\sum_{l\neq i}
(a_i^\dagger -a_l^\dagger )^{-1}
(\xi_{2i} -\xi_{2l} )K_{il}] -(1\leftrightarrow 2 )
\nonumber\\
&=&\nu \sum_{l\neq i}  (\xi_{1i}\prime-\xi_{1l}^\prime)
(a_i^\dagger -a_l^\dagger )^{-1}
(\xi_{2i} - \xi_{2l} )K_{il}
-(1\leftrightarrow 2 )\,,
\end{eqnarray}
which also vanishes because the summand is antisymmetric under the
interchange $i\leftrightarrow l$.

Finally, for
$X^{(2)}$ one gets from (\ref{eq:com1})

\begin{equation}
X^{(2)} =
\nu^2 \sum_{i,l,n} [(a^\dagger_i -a^\dagger_l )^{-1}(\xi_{1i} - \xi_{1l} )
K_{il},
(a^\dagger_i -a^\dagger_n )^{-1}(\xi_{2i} - \xi_{2n} )K_{in}]\,.
\end{equation}
The contribution to the summation of the terms with $l = n$ vanish
so that one is left
with

\begin{eqnarray}
X^{(2)}&&=
\nu^2 \sum_{i\neq l,i\neq n,l\neq n}\biggl (
(a^\dagger_i -a^\dagger_l )^{-1}(\xi_{1i} - \xi_{1l} )
(a^\dagger_l -a^\dagger_n )^{-1}(\xi_{2l} - \xi_{2n} )K_{il} K_{in}\nonumber\\
&&-
(a^\dagger_i -a^\dagger_n )^{-1}(\xi_{2i} - \xi_{2n} )
(a^\dagger_n -a^\dagger_l )^{-1}(\xi_{1n} - \xi_{1l} )K_{in} K_{il}\biggr )\,.
\end{eqnarray}
After replacing $l\leftrightarrow n$ in the second term we can
write $X^{(2)}$ as

\begin{eqnarray}
X^{(2)}&=&
\nu^2 \sum_{i\neq l,i\neq n,l\neq n}
(a^\dagger_i -a^\dagger_l )^{-1}
(a^\dagger_l -a^\dagger_n )^{-1}
(a^\dagger_i -a^\dagger_n )^{-1}\nonumber\\
&&\{(a^\dagger_i -a^\dagger_n )(\xi_{1i}\xi_{2l} +\xi_{1l}\xi_{2n} -
\xi_{1i}\xi_{2n} )
-1 \leftrightarrow 2\}K_{il}K_{in}\,.
\end{eqnarray}
Finally, using the cyclic property (2) of the elementary permutation
generators one easily verifies by permuting the summation indices $i,j,l$
that the $(a_i^\dagger - a_n^\dagger)$ coefficient in the second
line can be replaced by
$1/3[
(a_i^\dagger - a_n^\dagger) +
(a_l^\dagger - a_i^\dagger) +
(a_n^\dagger - a_l^\dagger)]$ which is identically zero.
This concludes the proof of the identity (\ref{eq:id4}).

In addition to the identities proven above
there also exist the following useful identities which can be
proven analogously
\begin{equation}
\label{eq:id1}
\sum_{i=1}^N \left \{ [ a_i, (\xi_1 (a_j^\dagger )], \xi_2
(a_i^\dagger )\right
\} = 2\xi_2  (a_j^\dagger ) \xi^\prime_1 (a_j^\dagger)\,,
\end{equation}
\begin{equation}
\label{eq:id2}
\sum_{i,j=1}^N \left [ [a_i, \xi_1 (a_j^\dagger ) ], [a_j,
\xi_2 (a_i^\dagger ) ] \right ] = 0\,.
\end{equation}

\section*{Appendix B:
The sketch of calculation of $[L_\xi, Q_\epsilon]$}

Below we outline the main steps of our calculation
of the $[L_\xi, Q_\epsilon]$ commutator for the case with $\lambda = 0$.

It is convenient to work with the following form for the Ansatz of
the $L_\xi$-generator which can easily be obtained from (\ref{svir})
with the aid of the formula (\ref{bsd}):

\begin{equation}
\label{LL}
L_\xi = \sum_i \xi_i D_i^{\theta,0} + {1\over 2}({\cal D}_i^0\xi_i)
{\cal D}_i
+\nu^2 \sum_{j\ne k \ne i}
\xi_i y_{ij}K_{ij}^{\rm tot} y_{jk}K_{jk}^{\rm tot}\,,
\end{equation}
with
\begin{equation}
D_i^{\theta,0} \equiv {\partial\over \partial z_i} + \nu\sum_{j\ne i}
{1\over z_{ij}}(1-K_{ij}^{\rm tot})\,.
\end{equation}

As a first stage we find for the commutator:

\begin{eqnarray}
\label{eq:comma}
&&
[L_\xi, Q_\epsilon] =
\sum_{i,j}\bigg (
\xi_i[D_i^{\theta,0},\epsilon_j]{\cal D}_j + \xi_i\epsilon_j
[D_i^{\theta,0},{\cal D}_j]
-\epsilon_i[{\cal D}_i,\xi_j]D_j^{\theta,0}\nonumber\\
&&+{1\over 2} ({\cal D}_i^0\xi_i)[{\cal D}_i,\epsilon_j]{\cal D}_j
+{1\over 2}\epsilon_j({\cal D}_i^0\xi_i)\{{\cal D}_i, {\cal D}_j\}
- {1\over 2}\epsilon_j\{{\cal D}_j, ({\cal D}_i^0\xi_i)\}{\cal D}_i
\bigg )\nonumber\\
&&+\nu^2 \sum_{i\ne j\ne k} \sum_l\bigg [\xi_i y_{ij}
K_{ij}^{\rm tot}
y_{jk} K_{jk}^{\rm tot}, \epsilon_l {\cal D}_l
\bigg ]\,.
\end{eqnarray}

We first consider the terms in the lowest order in $\nu$ which are
of the form $\xi\epsilon^{\prime} {\cal D}_i, \epsilon\xi^\prime
{\cal D}_i, ({\cal D}_i^0\xi_i)({\cal D}_i^0\epsilon_i){\cal D}_i$
and $\epsilon_i({\cal D}_i^0\xi_i)
D_i^{\theta,0}$. The first three type of terms gives us the
required supersymmetry transformation with the parameter (\ref{ps}).
%\begin{equation}
%\label{eq:sp}
%\bar\epsilon = \xi\epsilon^\prime - {1\over 2}\epsilon\xi^\prime + {1\over 2}
%({\cal D}_i^0\xi_i)({\cal D}_i^0\epsilon_i)\,.
%\end{equation}
The last type of term cancels\footnote{
In doing the calculation there is no need to write out the derivatives
${\cal D}_i$ and $D_i^{\theta,0}$ in terms of a $\nu$-independent part
and terms linear in $\nu$. Therefore, strictly speaking, our calculation
also takes care of a class of $\nu^2$-dependent terms in the commutator.}.
We next consider the remaining terms linear in $\nu$. They only come
from the first and second line in (\ref{eq:comma}).
We observe that all $\nu$-dependent terms in the second line of
(\ref{eq:comma})
are proportional to $({\cal D}_i^0\xi_i),$ whereas all terms in the
first line are proportional to $\xi_i$. This implies that the remaining
$\nu$-dependent terms occurring in the first and second line
should cancel independently. The second line in (\ref{eq:comma}) gives
rise to terms of the form\footnote{
We only give the general structure of the terms. We use a notation
where the specific name of the indices is irrelevant.}

\begin{equation}
\nu{\theta_{ij}\over z_{ij}}({\cal D}_i^0\xi_i)
\epsilon_iK_{ij}^{\rm tot}{\cal D}_i\,.
\end{equation}
These terms can be shown to cancel amongst each other without the need
to write out the extra $\nu^2$-dependent terms occurring inside ${\cal D}_i$
explicitly. Therefore,
this cancellation also involves a class of terms quadratic
in $\nu$.

We next consider the terms linear in $\nu$ coming from the first line
in (\ref{eq:comma}). The following three types
of terms linear in $\nu$ occur:

\begin{equation}
\nu{\theta_{ij}\over (z_{ij})^2}\xi_i\epsilon_j
(1-K_{ij}^{\rm tot}), \quad
\nu{\theta_{ij}\over z_{ij}}\xi_i\epsilon_j K_{ij}^{\rm tot}
{\partial\over \partial z_i},
\quad
\nu {1\over z_{ij}}\xi_i\epsilon_jK_{ij}^{\rm tot}{\cal D}_i^0\,.
\end{equation}
All three type of terms can be shown to cancel amongst each other.

We now consider the terms quadratic in $\nu$. In the second line of
(\ref{eq:comma}) only the anticommutator in the second term leads
to a $\nu^2$-dependent term which has not yet been considered. This
term cancels against a similar $({\cal D}_i^0\xi_i)$-dependent term
coming from the commutator in the third line of (\ref{eq:comma}).
This finishes the discussion of the second line in (\ref{eq:comma}):
all terms have either been cancelled or contribute to the supersymmetry
parameter $\bar \epsilon$ given in (\ref{ps}).
All remaining $\nu^2$-dependent terms come from the first and second line in
(\ref{eq:comma}). The following two type of terms occur:

\begin{equation}
\nu^2{\theta_{ij}\over (z_{ij})^2}\xi_i\epsilon_jK_{ij}^{\rm tot},
\quad
\nu^2 {\theta_{ij}\over (z_{ij})^2}\xi_i\epsilon_j K_{ij}^{\rm tot}
K_{jk}^{\rm tot}\,.
\end{equation}
The terms linear in $K_{ij}^{\rm tot}$  only come from the first line
in (\ref{eq:comma}) and it can be shown that they cancel amongst
each other. We next consider the terms quadratic in $K_{ij}^{\rm tot}$.
Once we have moved the permutation operators to the right we may distinguish
between two type of terms which should cancel independently of each other.
The first type of terms are proportional to $\xi_i\epsilon_i$,
i.e.~$\xi$ and $\epsilon$ have the same index; the second type of terms
are proportional to $\xi_i\epsilon_j$ with $i\ne j$.
Both type of terms can be further subdivided into either terms
with a summation over three  different indices $i\ne j\ne k$
or terms with a summation over two different indices $i\ne j$. A
straightforward but somewhat tedious calculation shows that all
type of terms cancel amongst each other.

We finally consider the terms trilinear in $\nu$. They only come from the
commutator in the third line of ({\ref{eq:comma}) and are given by

\begin{equation}
\nu^3 \sum_{i\ne j\ne k}\sum_{l\ne m}\bigg [
\epsilon_l(y_{lm}K_{lm}^{\rm tot}), \xi_i(y_{ij}K_{ij}^{\rm tot})
(y_{jk}K_{jk}^{\rm tot})\bigg ]\,.
\end{equation}
We first consider the case that either $l$ or $m$ is equal to $i,j$ or $k$
but not both. The terms in the commutator corresponding to this
case can always be written into the following three type of
terms\footnote{The specific name of the indices in this formula
is important now.}:

\begin{equation}
\nu^3
\sum_{i\ne j\ne k\ne l} \bigg (\epsilon_i\xi_i \ y_{ij}y_{ik}y_{il} \times
K^3\,, \hskip .4truecm
\epsilon_i\xi_j\ y_{ij}y_{jk}y_{jl}\times K^3\,,
\hskip .4truecm
\epsilon_i\xi_j\ y_{ij}y_{jk}y_{il}\times K^3 \bigg )\,.
\end{equation}
To do this one must make use of the $y$-identities given in
(\ref{eq:y-id})
A tedious calculation shows that all three type of terms cancel.
Finally, we consider the case that both $l$ and $m$ are equal to
$i,j$ or $k$. It can be easily shown that this case always leads
to terms which identically vanish.

It is straightforward to extend the calculation of the $[L_\xi,
Q_\epsilon]$
commutator to non-zero values of $\lambda$.

\end{document}